# Fusing restricted information

Magnus Jändel, Pontus Svenson, Ronnie Johansson
Division of Information and Aeronautical Systems
FOI Swedish Defence Research Agency
Stockholm, Sweden
firstname.lastname@foi.se

*Abstract*—Information fusion deals with the integration and merging of data and information from multiple (heterogeneous) sources. In many cases, the information that needs to be fused has security classification. The result of the fusion process is then by necessity restricted with the strictest information security classification of the inputs. This has severe drawbacks and limits the possible dissemination of the fusion results. It leads to decreased situational awareness: the organization knows information that would enable a better situation picture, but since parts of the information is restricted, it is not possible to distribute the most correct situational information.

In this paper, we take steps towards defining fusion and data mining processes that can be used even when all the underlying data that was used cannot be disseminated. The method we propose here could be used to produce a classifier where all the sensitive information has been removed and where it can be shown that an antagonist cannot even in principle obtain knowledge about the classified information by using the classifier or situation picture.

## I. Introduction

Information fusion deals with the integration and merging of data and information from multiple (heterogeneous) sources. Most fusion research deals with new algorithms for merging information or describes systems that solve fusion problems in a specific setting. Fusion is about helping decision makers or analysts increase their situational awareness by reducing uncertainty or information content.

In many cases, the information that needs to be fused has a security classification. The result of the fusion process is then by necessity restricted with the strictest information security classification of the inputs. This has severe drawbacks and limits the possible dissemination of the fusion results. It leads to decreased situational awareness: the organization knows information that would enable a better situation picture, but since parts of the information is restricted, it is not possible to distribute the most correct situational information. Instead, many users must use situation information that the organization knows to be incorrect.

Some example application use cases where it is necessary to fuse restricted information are:

- intelligence analysis, where for instance concerns about the security of a source might lead to the situational picture not being updated. For instance, it might be the case that a highly placed human source gives detailed information (referred to as HUMINT for human intelligence) about, e.g., the order of battle that when combined with input from sensors could be used to determine what type of units are being deployed in a foreign country. However, the fusion system available to country analysts is not able to use the HUMINT information, since that could potentially reveal the source. All the information is only available to, e.g., the commander in chief. By developing methods for secrecy-preserving information fusion, it could be possible to at least give some information to the low level commanders, with a guarantee that this will not reveal the source.

- classification and screening tasks, where it, for ethical and privacy reasons, is not permitted to give all users of a classification algorithm detailed knowledge about the data. In a peace-keeping mission it might for example be of interest to fuse military intelligence data with local police records and a database of registered voters in a recent general election. The voter registration data includes, however, religious affiliation which local authorities, for ethical reasons, are reluctant to share with the peace-keeping force.

- there might be restrictions in what kind of information can be sent over a communication line or given to an expeditionary force. Concerns about the integrity of the communication lines and fear of losing databases to attacks today prevent some information from being shared with national intelligence cells in countries where a force is conducting a peace-enforcing operation. Secrecy-preserving fusion would make it possible to share more information with local intelligence cells.

Figure 1 below attempts to show the fusion use case. In this simplified example, one user (right) has access to all three sources and can confirm what kind of object is being seen. The other user (left) is not allowed access to the sensitive signature database at all and thus cannot identify a certain object. In this paper, we take a first step towards developing a method for allowing the user to get more fine-grained access to the data in the sensitive database, thus enabling better identification.

The approach could be compared to object-based security. In Object-based security, the classification of a document determines who can access it, not the database in which it is stored. This is however not always a possible solution.

We note that the exact reason why some of the information is secret can vary. The information might come from a source whose identity needs to be protected. The information might come from a new sensor system whose detailed capabilities must be protected. The information might reveal weaknesses in our capabilities. Our suggested method is agnostic to the specific reason for secrecy.

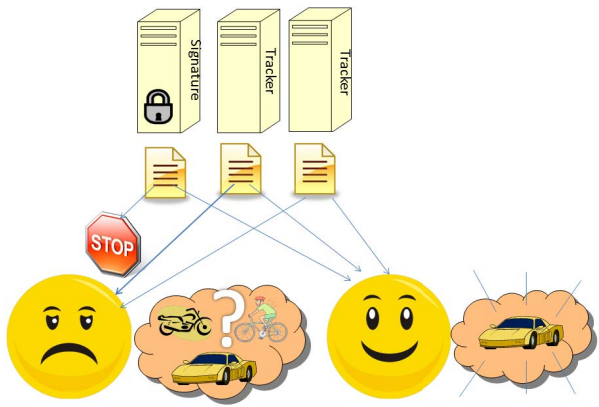

Figure 1 – Toy example showing two users with different access rights to databases and sources. There are two tracking sensor systems that output position and signature information. A classifier uses the signature from the sensors and a signature database to classify the object. In this case, some of the signature database information is restricted. The happy user (right) has a high clearance and is allowed access to the full classifier, so can determine that the object seen is a sports car. The unhappy user (left) is not permitted access to the signature database, and is thus completely ignorant of the type of the object. In this paper, we take some first steps towards building a secrecy-preserving classifier. The secrecy-preserving classifier is designed to use some of the information in the classified database, but comes with guarantees that it is not possible to use it to reveal the secret information in the database.

In this paper, we take steps towards defining fusion and data mining processes that can be used even when all the underlying data that was used cannot be disseminated. The method we propose here could be used to produce a classifier where all the sensitive information has been removed and where it can be shown that an antagonist cannot even in principle obtain knowledge about the classified information by using the classifier or situation picture.

This paper is outlined in the following way. First, we describe some related work. We then introduce a formal way of describing fusion processes and the sensitivity constraints on the information. Next, we apply automata theory to filtering the sensitive information and discuss how to optimize the fusion process.

### A. RELATED WORK

As far as we know, few works or none on fusion of restricted sources exist to date. For privacy preserving information processing, however, Information Fusion can be used to support a data mining task. [1] presents two different approaches for how to use information fusion to preserve privacy in data: Data aggregation and record linkage. Data aggregation is basically a descriptive modelling task. The idea behind the data aggregation approach is to cluster data (based on some similarity measure) and then for each cluster (aggregate) to fuse its members to find a resulting representative of the cluster. Various fusion operators have been proposed, and the selection of operator is typically dependent on the type of the data. The purpose of record linkage is to match data from different databases which refer to the same entity. Privacy preserving record linkage, then, concerns drawing conclusions based on the linked data, without revealing the discovered links. The fusion step of the process involves aggregating the selected (linked) data.

## II. BACKGROUND

In this section we give a formal description of the problem we are focusing on. We are working with a simplified case where the goal is to learn a classifier based on partly restricted information. We are assuming that there is a database that is used to learn the classifier, and that the opponent has access to some additional information. The goal of the opponent is to use the classifier to reveal the restricted information in the database.

The generalization to fusion processes is straightforward: what we are studying here is not the situation picture per se, but rather the model that is used by commanders to create their situation picture. The classifier represents a fusion algorithm that has background or doctrine knowledge built-in. In the case where a low-level commander needs access to some restricted information, that information would be built into the fusion model by the high-level command and provided to the low-level commander. The high-level command ensures that the fusion model contains all background knowledge/information that is necessary for the low-level commander, but also makes the fusion model weak enough so that no sensitive information can be revealed by using it. The low-level commander then fuses the data and information available to them.

Our problem can be succinctly described as finding a classifier that is strong enough to be useful, but weak enough so that it does not reveal any secret information.

### A. Describing the fusion process formally

A classifier is a function $f$ that maps a set of attribute values $A$ to a class $c$, i.e., $f(A) = c$. In a generalization of $f$, not only a single preferred class is returned, but a degree of confidence (conventionally expressed as a probability) is assigned to all possible classes, i.e., $f(c;A)=p(c)$. The classifier also implicitly depends on the data available to build it.

Building a classifier can be interpreted as first fusing data (off-line) into a resulting model of the data, which can subsequently be used as a fusion operator that can fuse sensed attribute values (on-line).

The process for building a classifier then consists of the following main elements:

1. Build a classifier
2. Filter out sensitive information (this step might include building new classifiers on subsets of the available information or modifying the classifier from step 1 so that no sensitive information is revealed)
3. Fuse information

where the last element only applies if we have several source databases to consider. Different permutations of these process elements make for different algorithmic problems to consider. Restricting the discussion to one single database we have the permutations Filter →Build and Build→Filter, where the former was briefly discussed in the previous sub-section. With several databases and taking into account that the same process does not have to be applied to all databases, we have a large number of possible permutations as illustrated in Figure 2 in which we have applied the rules that fusion, filtering and building is applied only once to a given artifact. Even with this somewhat limiting assumption we note that there is a plethora of possible algorithm architectures even if we only consider a few databases.

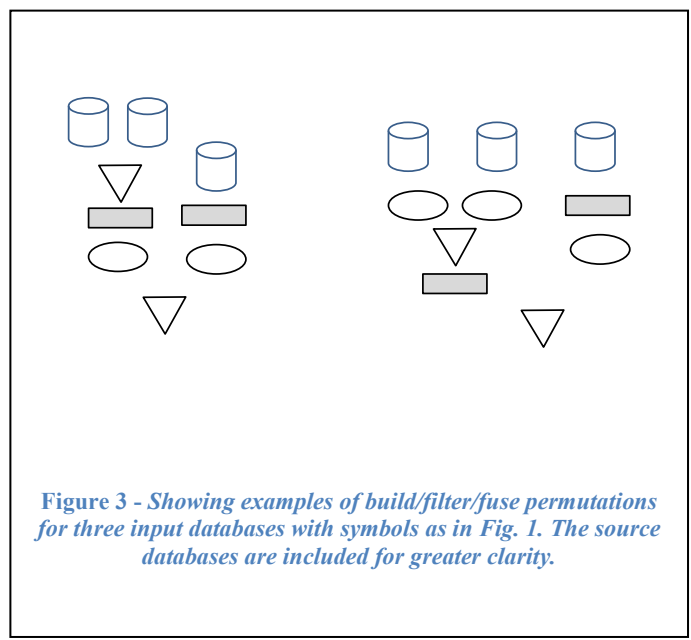

Figure 3 - *Showing examples of build/filter/fuse permutations for three input databases with symbols as in Fig. 1. The source databases are included for greater clarity.*

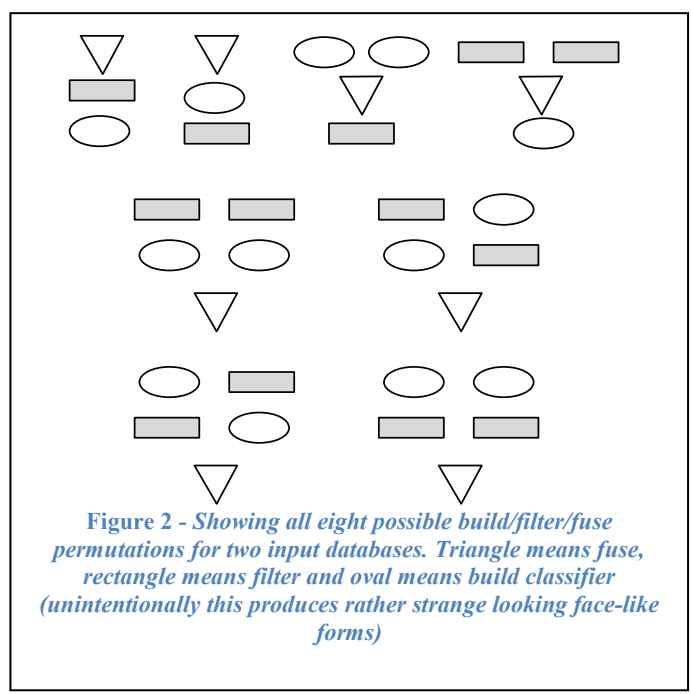

Figure 2 - *Showing all eight possible build/filter/fuse permutations for two input databases. Triangle means fuse, rectangle means filter and oval means build classifier (unintentionally this produces rather strange looking face-like forms)*

In the case of three databases, we encounter processes corresponding to more complex permutations some of which are illustrated in Figure 3.

To enable easy reference to the various configurations in Figure 2 and Figure 3, we need a language similar to chemical formulas for describing molecules. Our *atoms* are the operations build (B), filter (E) and fuse (F) where we use the symbol E for filter to avoid symbol collisions and with the mnemonic that filtering is a form of editing. Note that the filtering operator E is overloaded since it can apply to a database or a classifier depending on the context. Each process formula describes the operators that are applied to the databases in temporal order from the right to the left in the formula using the function composition symbol ∘ to separate operators that are applied to the same database and the symbol $^\times$ to separate groups of operators that are applied to different databases. Hence ∘ indicates serial processing while $^\times$ means parallel processing. In this notation the two options for processing a single database is E∘B and B∘E where the former indicates that we first build a classifier and subsequently filter the restricted information from the classifier. The eight options for processing two databases (see Figure 2 in the order left to right and top to bottom) B∘E∘F, E∘B∘F, E∘F∘(B×B), B∘F∘(E×E), F∘(B×B)∘(E×E), F∘(B×E)∘(E×B), F∘(E×B)∘(B×E) and F∘(E×E)∘(B×B). Similarly we can write the two configurations in Figure 3 (right to left) as F∘(B∘E∘F) × (B∘E) and F∘(E∘F∘(B×B)) × (B∘E) respectively.

If we want to express how the processes are applied to specific databases we can specify the sensitivity descriptors that the processes operate on. For example, F∘(E×B)∘(B×E){$DB_1$; $DB_2$} , where we apply the convention that parallel operators are applied in left-right order to the arguments, means that $DB_2$ is filtered before the classifier is built while the opposite applies to $DB_1$.

Note that this notation is not restricted to only being used for describing sensitive operations. It can be used to define any distributed fusion process. For instance, Bowman has introduced the Dual node network (DNN) to decompose complex fusion processes into a network of interconnected fusion nodes [2]. The approach we propose here could be used also to describe DNNs.

While the discussion in this paper focuses on classifiers, the extension to fusion algorithms in general is straightforward. A sensor can be seen as a database [3]. The build step corresponds to constructing situation information. The resulting situation information elements are then used to answer queries from the decision makers. The confidential information that we wish to preserve must be hidden from these users – they must not be able to infer restricted information from the situation picture.

A concrete example could be that we have access to a new, advanced sensor that is able to collect multi-spectral information. The fact that we have such a sensor is classified and only high-level commanders are allowed to see the information from it. The secrecy-preserving problem is to determine what results from the sensor that we can fuse with other information and send to low-level commanders, without them (or an adversary) being able to infer that we have access to the new sensor. A more traditional example is that we have a human source whose existence must not be revealed. The secrecy-preserving fusion problem is to fuse the information from the secret agent with other sources and then weaken the situation picture information enough so that the source is not revealed.

III. PROBLEM FORMULATION

A database (DB) includes both *sensitive information* (SI) and *open information* (OI). What is sensitive and not depends on the objectives and knowledge of the adversary and on what kind of security breaches that will be realized if the adversary gains access to the sensitive data. In the following we will always keep in mind that the term *sensitive data* is always construed in the context of a specific adversarial scenario.

In our example with the peace-keeping force that wants to fuse intelligence data with local police records and a voting database, consider for example that the de-identified voting database is turned over to the police to be fused with police information, filtered and then passed on to the peace-keeping force. A local police officer could easily pin-point the exact identity of a villager based only on religious affiliation, gender and age while a peace-keeping intelligence officer without local information on who is practicing what religion would be unable to re-identify the same person.

A. Stakeholders and trust relations

There are four types of stakeholders,

1) The database owner (DBO)
2) The classifier producer (CLP)
3) The Certifier (CTF)
4) The user of the classifier (CLU)

The CLP builds the classifier using data provided by the DBO. The CTF represents the primary stakeholders in the integrity of the sensitive data and approves or rejects the release of the classifier to the CLU.

There are three essential pieces of information that the stakeholders may be trusted or not trusted to have access to,

1) The sensitive information in the database (SI)
2) The source code of the classifier (CL-WB, for classifier white-box)
3) The object code of the classifier (CL-BB, for classifier black-box)

We realize that the definition of object code is not clear-cut but use this term under the simplifying assumption that the user only has access to the input and outputs of object code but not to its internal structure. The source code is a white box and the object code is a black box view of the classifier. Having access to the source code including the learning state of the classifier is more or less useful depending on the type of algorithm employed.

Table 1 maps out the possible trust relations between the respect to the four stakeholders and the three information artifacts. We assume that all stakeholders are trusted with the open data. In Table 1, the symbol T means that the stakeholder must be trusted with the information object whereas the blank field means that we have choice on whether or not to trust the stakeholder with the information.

Table 1. Trust relations for different stakeholders. See text.

| Stakeholder | SI | CL-BB | CL-WB |
|---|---|---|---|
| DBO | T | | |
| CLP | | T | T |
| CTF | T | T | T |
| CLU | | T | |

(Note: any stakeholder who has access to the CL-WB also has access to the CL-BB. However, the converse is not true. It might also be the case that there are several layers of SI, where the CLU has access to some of them but not all.)

Different technical solutions apply depending on how we fill in the blank fields of table 1. Consider for example a case in which the CLP is not trusted with the sensitive information. In that case the only option will be to let the DBO extract the open information and provide to the CLP. The CTF should verify that the filtered database in fact does not include any sensitive data but there is no need to certify the classifier since it only is based on the open information. This scenario is strongly related to the task of anonymizing a database according to techniques described in [4]. A systematic approach to the certification process is described in [5].

Things get more interesting if we consider the case where the classifier producer (CLP) is allowed access to the sensitive information but the user of the classifier (CLU) is not. This is the main motivating example for the method proposed in the paper. In this case, the producer (CLP) must make sure that all remnants of sensitive information (SI) is removed from the classifier before it is submitted to the CLU for use. The certifier must verify this, and hence also requires access to the SI. This use case directly relates to the motivating examples given in section I and at the end of section III. In all these examples, the user of the fusion system (classifier) is not allowed un-restricted access to all information. Hence there is a need to make the classifier/fusion system weaker than it could be, in order to ensure that no secret information can be revealed by applying it to new data.

*B. Sensitivity descriptors*

Sensitive information may be isolated to one database or depend on correlations between several databases. To compactly capture the sensitivity situation we introduce *sensitivity descriptors* where the relevant databases and their sensitivity relations are listed. Sensitivity descriptor $S_1=\{DB_1; DB_2\}$ means for example that both databases have sensitive information but that the sensitive information is uncorrelated between the two databases. The descriptor $S_2=\{DB_1;[DB_1, DB_2]\}$ means that $DB_1$ has individual sensitive information and that there also is sensitive information that depends on correlations between $DB_1$ and $DB_2$. A sensitivity descriptor consists in general of a series of correlation descriptors separated by semicolons. Each correlation descriptor is a set of database symbols separated by commas indicating that sensitive information depend on correlations between the databases in the list.

We are now in the position to compactly describe what processes that work for which sensitivity configurations. As an example, consider the sensitivity descriptor $S_3=\{[DB_1, DB_2]\}$. We note that the correlated sensitivity can only be handled if we fuse before we filter.

So the only workable processing configurations are:
- B∘E∘F,
- E∘B∘F,
- E∘F∘(B×B)

The other possibilities are not allowed:
- B∘F∘(E×E),
- F∘(B×B)∘(E×E),
- F∘(B×E)∘(E×B),
- F∘(E×B)∘(B×E)
- F∘(E×E)∘(B×B)

The process by which we derived the workable processing configurations can be automated. Given a set of candidate configurations and the sensitivity descriptor, the configurations can be filtered until only the allowed ones remain.

For more complex sensitivity configurations we can apply similar rules. It is for example easy to see that the sensitivity situation described by descriptor $S_4=\{[DB_1, DB_2]; DB_3\}$ is handled by the process F∘(B∘E∘F)×(E∘B).

IV. FILTERING OF SENSITIVE INFORMATION

We now proceed to describe a way to actually generate the allowed fusion processes. We will make use of formal languages and automata theory for doing this.

*A. Automata theory and grammars as a way of delimiting the allowed fusion structures*

The problem we are looking at in this section is: can we find a succinct characterization of the allowed fusion processed that takes the form of an automaton/language? We solve this by using context-sensitive grammars [6].

If we are able to further characterize desirable fusion structures in terms of additional constraints on the language generated, we could exploit this in two ways
1. generating all possible allowed fusion structures in a simple way by simply applying the production rules. This would also give us a distance in the space of all allowed fusion structures (the distance measure would be the hierarchical distance between two structures: the distance to the latest common ancestor word)
2. possibly use techniques from automata theory to analyse different fusion structures and determine what additional properties they would have.

This approach is not limited to the current problem but could actually be a general method for analyzing distributed fusion systems. In addition to the security properties, also issues relating to information pedigree and credibility and reliability of the information could be analyzed in this way.

*B. Filtering schemas*

As preparation for defining a formal method for transforming a sensitivity descriptor to a process finite automata, we define filtering schemas for describing the temporal order of the filtering operations. The filtering schemas will serve as skeletons for building finite automata.

To be more specific about the effect of a filtering operation we introduce an argument that specifies what databases that are effected and what kind of sensitivities that are removed by the operation. The argument employs the same kind of notation as in the sensitivity descriptor but refers only to indices of databases. The filter E(1) removes the sensitive information from the database $DB_1$. The filter E([1,2]) removes the correlated sensitive information from databases $DB_1$ and $DB_2$ but not uncorrelated sensitive information from any of $DB_1$ and $DB_2$. A filtering schema is a sequence of filtering operations as for example E(1) E([1,2]) meaning that we first remove the correlated sensitive information from $DB_1$ and $DB_2$ and subsequently remove the sensitive information from $DB_1$. We can merge filtering operators as for example in E(1) E([1,2]) →E(1;[1,2]). The arrow indicates transformation from one schema to another (and not equality). The schema E(1;[1,2]) means that the individual sensitivity of $DB_1$ is removed in the same atomic filtering process as the correlated sensitivity in $DB_1$ and $DB_2$.

The information may be removed from the databases or from classifiers that are trained on the databases. The filtering schema is just a plan for the temporal order of the filtering and does not specify the targets of the filtering operations or any other details of the process.

Using this extended notation for filtering operations we can now define a simple algebra for the temporal ordering of filtering that is best described by a few examples.

$E(2)E(1)\{DB_1; DB_2\} \rightarrow E(2)\{DB_2\} \rightarrow \{\}$.
$E([1,2])E(1)\{DB_1; [DB_1, DB_2]\} \rightarrow E([1,2])\{[DB_1, DB_2]\} \rightarrow \{\}$.
$E(1)E([1,2])\{DB_1; [DB_1, DB_2]\} \rightarrow E(1)\{DB_1\} \rightarrow \{\}$.
$E(1, [1,2])\{DB_1; [DB_1, DB_2]\} \rightarrow \{\}$.

A sequence of filters that transforms a sensitivity descriptor to the null descriptor $\{\}$ is a filtering schema of the descriptor.

### C. Generating sensitivity descriptors using context sensitive grammars

In this section we give an example of how the sensitivity descriptors can be generated using context sensitive grammars (CSG) [6]. We first look at the case of two databases.

#### 1) Two databases

For two databases there are eight sensitivity descriptors $S_1=\{\}$, $S_2=\{DB_1\}$, $S_3=\{DB_2\}$, $S_4=\{DB_1; DB_2\}$, $S_5=\{[DB_1, DB_2]\}$, $S_6=\{DB_1; [DB_1, DB_2]\}$, $S_7=\{DB_2;[DB_1, DB_2]\}$ and $S_8=\{DB_1; DB_2;[DB_1, DB_2]\}$ where $S_1$ means that there are no sensitive information in any of the databases $S_2$ means that there are sensitive information in $DB_1$ but not in $DB_2$ and $S_4$ means that there are sensitive information in both $DB_1$ and $DB_2$ but that there are now correlation between the sensitivities of the databases.

In this section we define a CSG that generates these sensitivity descriptors and nothing else. In the resulting language we want to avoid duplicates due to permutations and get the symbols in lexicographic order. The CSG is defined as follows.

The variables are $^1d_1$, $^1d_2$, $^2d_1$
The terminals are $DB_1$, $DB_2$, $\{$, $\}$, $[$, $]$, $;$
$S$ is the start symbol

The production rules are,
  $S \rightarrow \{\}$
  $\{\} \rightarrow \{^kd_i\}$ where $k=1,2$ and $i=1,2$
  $^kd_i\} \rightarrow {}^kd_i; {}^{k'}d_{i'} \}$ where $k' \geq k$ and $i' > i$ if $k'=k$
  $^1d_i \rightarrow DB_i$
  $^2d_1 \rightarrow [DB_1, DB_2]$

#### 2) Arbitrary number of databases m

For the set of all possible duplicate-free and lexicographically ordered sensitivity descriptors for m databases is generated by the following CSG.

The variables are all symbols $^kd_i$ where $k=1, 2, \ldots m$ and $i = 1,2, \ldots A_{mk}$ in which $A_{mm} = 1$ and $A_{mk} = m(m-1)(m-2)\ldots(k+1)$.
The terminals are $DB_1$, $DB_2$, $\ldots DB_m$, $\{$, $\}$, $[$, $]$, $;$ .
$S$ is the start symbol.

The production rules are,
  $S \rightarrow \{\}$
  $\{\} \rightarrow \{^kd_i\}$ where $k$ and $i$ take any allowed values
  $^kd_i\} \rightarrow {}^kd_i; {}^{k'}d_{i'} \}$ where $k' \geq k$ and $i' > i$ if $k'=k$
  $^1d_i \rightarrow DB_i$
  $^2d_i \rightarrow [DB_a, DB_b]$ where $a$ and $b$ are defined by some lexicographic order over the $A_{m2}$ possible sensitivity descriptors with two databases.
  $^kd_i \rightarrow [DB_a, DB_b, \ldots DB_q]$ where $k$ indexes $a$, $b$, ... $q$ are defined by some lexicographic order over the $A_{mk}$ possible sensitivity descriptors with $k$ databases.

### D. Building process finite automata from filtering schemas

Given a filtering schemas for a given sensitivity descriptor we are now ready for fleshing out the details of the related finite automaton. Figure 4 provides an example of how the filtering schema $E(1)E([1,2])$ for the sensitivity descriptor $\{DB_1; [DB_1, DB_2]\}$ translates to a finite automaton. We note that the automaton is modular with one component for each filtering operator. In Figure 4 we introduce a null action T for marking the transition from the clean final state of one component to the initial state of the next component.

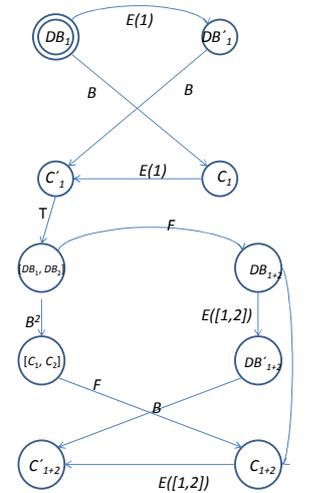

**Figure 4** - *The processing finite automaton based on the filtering schema E(1)E([1,2]). Note that the initial state is $DB_1$ because E(1) comes first in the filtering schema.*

In general we can build any process finite automaton by matching the filter operators of the filtering schema to automata components in a process library.

## V. ACTOR TRUST AND COMPETENCE CONSTRAINTS

Once we have found the finite automaton that summarizes the abstract states and processing steps of the information fusion process we are in the position to analyse trust and competence factors related to the various actors and stakeholders that are involved in the process.

The finite automaton is useful for the trust and competence analysis since the states refers to information which we will have to trust some actor to safeguard and make use of. The transitions of the finite automaton are the key processing steps for which we need to find actors that are competent to perform the steps. Therefor we will now describe a process for handling trust and competence constraints that is based on the finite automaton view of the problem and the solution space. The finite automaton in Figure 5 which is based on the sensitivity descriptor $\{[DB_1, DB_2]\}$ will serve as a running example for this section. The analysis phases are as follows.

Firstly, identify the finite automaton states that are sensitive. In Figure 5 those states are marked with red color. The start state is not marked as sensitive since it is a prerequisite that a database owner has access to the sensitive data in the owned database. The fused database $DB_{1+2}$ and the fused classifier $C_{1+2}$ are marked since they contain the sensitive correlations between the component databases. The state $[C_1, C_2]$ might also be sensitive since the two different actors that have access to each of the component classifiers jointly and deceitfully could get access to the sensitive information. The states marked with a prime do only contain filtered information and are thus not sensitive.

Secondly, for each sensitive state list the actors that can be trusted with the information referenced by the state. For the state $DB_{1+2}$ of Figure 5 we assume that actors $A_1$, $A_2$ and $A_3$ are trusted. Sensitive states that have no related trusted actors are removed from the finite automaton schema.

Thirdly, for each remaining sensitive state, for each trusted actor related to the state and for each action linked to the state investigate if the actor is competent to perform the action. The resulted is summarized in a *competence table* related to each sensitive state. Next to the state $DB_{1+2}$ of Figure 5 we exemplify a competence table where T means that the actor is competent to perform the action and F means the opposite. Transitions that have no related trusted and competent actor are removed from the finite automaton and from the competence table. Hence we should remove action *B* from the state $DB_{1+2}$ in Figure 5. The output this far is *the trust and competence constrained finite automata*.

Fourthly, extract the accepted language from the trust and competence finite automata. This will be the set of all feasible processing options.

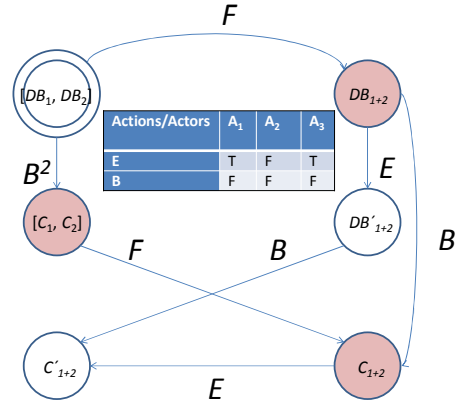

**Figure 5** - *Example of how to apply actor trust and competence constraint to a finite automaton. Sensitive data combinations are symbolized by nodes marked with red colour.*

## VI. PUTTING IT ALL TOGETHER

The overall procedure for analyzing and finding a solution in a situation that involves fusing information in databases containing restricted information is as follows.

*1) Define the sensitivity descriptor according to the procedure in section IV.C.*

*2) Define the set of applicable filtering schemas according to section IV.B and express this set as a finite automaton as described in section IV.D.*

*3) Analyse trust and competence constraints according to section V. This will result in a constrained finite automata that defines all feasible information processing solutions. The set of process descriptions that are in the allowed language of any of the constrained finite automata are in the feasible set of processes.*

*4) Select one of the feasible process descriptions as the preferred solution based on judgments including ethical, political, economical and other real-world factors not included in steps 1-3. And of course optimization of the fusion process*

The steps above ensure that the sensitivity/security aspects of the fusion process are taken care of. Of course, we also need to ensure that the fusion process is efficient so that the situational picture produced is accurate enough. This process consists of the following steps:

1. define a space of all allowed fusion processes
2. search this space using the sensitivity criterion as a constraint on allowed processes/structures and with a normal evaluation criterion for the efficienty and accuracy of the fusion process.

As an example, assume that we have a measure of efficiency of the fusion process. This could be the distance to the best classifier for the problem, or a comparison to ground truth for a general situation assessment system. For many problems, the users are not interested in the ideal solution – e.g., customs and police might only be interested in knowing if there is a higher than 80% likelihood that a passenger is a terrorist.

Let $F(K, \pi)$ be a fusion process where $K$ is classifier and $\pi$ is problem. Also assume that there is a risk functional $R(K;A)$ where $A = \{D, V, \chi\}$ and $D$ is antagonist's other available data, $V$ is goals and purpose of the antagonist, $\chi$ is the loss or damage caused by the antagonist knowing the sensitive information.

We learn the classifier $K$ from data. $K = K(data,$ machine learning algorithm, structure $S)$. By using different structures $S$ (including also the filtering rectangles), we get different classifiers $K$.

Now plot Risk versus $F$ in a diagram. Two possible problem formulations: we are interested in getting the best possible $F$ given that $R$ is less than $R^*$; we are interested in getting a $K$ with at least efficiency $F^*$ and want the risk R to be as low as possible given $F(K) > F^*$.

This optimization could also be formulated as search in a fitness landscape – where the ethical requirements prevent us from accessing parts of the landscape. Compare to linear programming where we want to find the best $K$ that lies on the boundary separating allowed structures from dis-allowed.

## VII. EXAMPLE

In this example, we have two databases, use support vector machines (SVM) as classifiers and make several simplifying assumptions that will be introduced as needed. Each record in both of the databases consists of a two-dimensional input vector **x** with real-valued elements and a binary label. The two databases have the same input vector in the same record but with different labels. The i:th records of database $DB_1$ and $DB_2$ are hence $[y_i, \mathbf{x}_i]$ and $[z_i, \mathbf{x}_i]$ respectively where the input vector $\mathbf{x}_i$ is the same in both records.

According to the first analysis step of section IV we generate the full spectrum of possible sensitive relations as the resulting language of the appropriate context sensitive grammar (see subsection IV.C where the relevant sensitivity relations are listed as an example). Furthermore we ponder the nature of the data and possible attack scenarios with the result that the appropriate sensitivity descriptor is selected to be $S = \{[DB_1, DB_2]\}$ meaning that the sensitivity of the information is related to correlations between data in both of the databases.

In the second step of section IV we summarize the filtering schemas (see subsection IV.B and IV.D) that in principle could remove the sensitive information as a finite automaton (see subsection IV.D). Figure 6 displays the output of this step.

The third step of section IV introduces trust and competence constraints with a result that is showed in Figure 5. We conclude that the feasible options are represented by the filtering schemas B∘E∘F and E∘F∘(B×B).

In the fourth step of section IV we judge that the most economical solution based on the capacity and cost profile of the involved organizations is the filtering schema E∘F∘(B×B) where each data owner first builds a support vector machine classifier and subsequently one of the organizations fuse and filter the classifiers.

Specializing the abstract filtering schema to our specific example with binary data and support vector machines, we start by building two binary SVM trained to classify $y$ and $z$ respectively. The two leftmost upper panels of Figure 6 show the data and the two leftmost lower panels show the resulting support vectors. Fusing the SVM means that the combined classifier forwards the test vector to each component SVM and outputs the combined results, namely $y$ and $z$.

Assume now that that the existence of a region where $y,z=1,1$ is the sensitive fact that we want to filter out (see the upper rightmost panel of Figure 6. There are several options for filtering. If the application allows us to make a black-box classifier we could just add some post processing logic after the dual SVM for example by outputting and $y,z=0,1$ if the result of the SVMs is and $y,z =1,1$. If we must produce a white-box classifier we could edit the support vectors of the component SVMs in our example according to the following procedure.

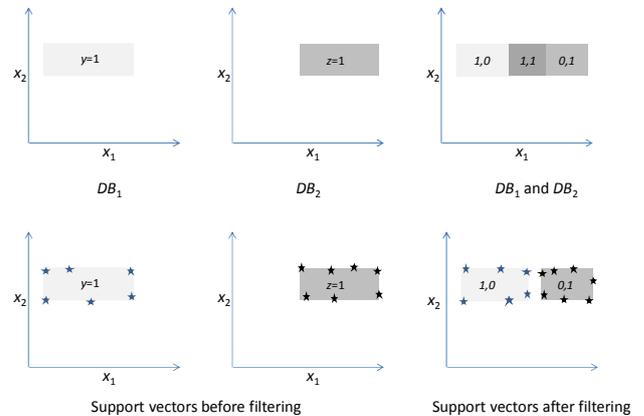

Figure 6 - *Example of* E∘F∘(B×B) *processing for a two-dimensional input vectors and binary labels. The upper left and middle panels show the domains where y=1 and z=1 respectively. The upper right panel shows the fused database with domains y,z=1,0, y,z=1,1 and y,z=0,1 respectively. We assume that the existence of a y,z=1,1 is the sensitive fact. The lower left panel and middle panels shows the support vectors (stars) before filtering and the lower right panel shows the filtered support vectors.*

For each support vector in the *y*-classifier that classifies as *z*=1 move it for a distance δ in opposite direction of the gradient of the classification function of the *z*-classifier. Repeat until all support vectors of the *y*-classifier is classified as *z*=0 by the z-classifier. This simple algorithm assumes a problem geometry as in Figure 6 and uses the fact that the initial support vectors ideally is on the convex hull of the *y*=1 domain. The resulting filtered support vectors are showed in the rightmost lower panel of Figure 6. After manipulating the support vectors we also need to adapt the SVM weights to the new support vector configuration. Since we lack access to the training data, this can be done by using the support vectors as training samples when retraining the weights. The fused and filtered classifier is now ready for certification.

## VIII. Discussion

In this paper, we discussed a method for weakening a classifier or fusion model so that it is not possible to use it to reveal sensitive or restricted information. This is important when fusing information from several data sources with different classification levels. The method also has applications in civilian security, where it for integrity and privacy reasons is necessary to ensure that a user of a classifier cannot use the classifier to reveal sensitive information about citizens.

Using a fusion method that allows accurate fusion of restricted information is very important in setting where compartmentalization of information is important. A drawback of normal fusion methods and the desire to go from "need to know" to "need to share" is that it enables a single infiltrator to potentially reveal all the classified information that is used to construct situation assessment. By moving to a paradigm where fusion models are constructed so that they cannot be used to reveal sensitive information, while still ensuring that they are strong enough so that they are useful, this problem could be solved.

In this paper, we took only a few first steps towards defining such a fusion process. We suggested a method for formalizing the sensitivity constraints of data sources and suggested a method for generating permitted fusion processed as strings in a language generated by a finite automaton.

In future work, we aim to test the method on traditional fusion data and extend the fusion process modelling method introduced to general distributed fusion processes.

### Acknowledgment

The research presented here was funded by the R&D programme of the Swedish Armed Forces and by the European Union under FP7 Grant Agreements 284862 (FIDELITY) and 261679 (CONTAIN).